\documentclass{JHEP3}
\long\def\symbolfootnote[#1]#2{\begingroup%
\def\thefootnote{\fnsymbol{footnote}}\footnote[#1]{#2}\endgroup} 

\usepackage{epsfig}

\newcommand{\be}{\begin{equation}}
\newcommand{\ee}{\end{equation}}
\newcommand{\tr}{{\rm Tr}}

\newcommand{\vpot}{{\mathit a}}
\newcommand{\LL}{{\cal L}}
\newcommand{\bea}{\begin{eqnarray}}
\newcommand{\eea}{\end{eqnarray}}
\newcommand{\Tr}{{\rm Tr}}

\author{R. Narayanan
\\Department of Physics, Florida International University, Miami,
FL 33199, USA\\E-mail: \email{rajamani.narayanan@fiu.edu}}
\author{ H. Neuberger
\\ Rutgers University, Department of Physics and Astronomy,
Piscataway, NJ 08855, USA\\E-mail: \email
{neuberg@physics.rutgers.edu} }

\title{Two dimensional fermions in three dimensional YM}

\abstract {Dirac fermions in the fundamental representation of $SU(N)$ 
live on the surface of a cylinder embedded in $R^3$ and 
interact with a three dimensional $SU(N)$ Yang Mills vector potential preserving
a global chiral symmetry at finite $N$. As the circumference of the cylinder 
is varied from small to large, 
the chiral symmetry gets spontaneously broken in the
infinite $N$ limit at a typical bulk scale. 
Replacing three dimensional YM by four dimensional YM introduces non-trivial 
renormalization effects. 
}


\preprint{}

\begin{document}

\section{Introduction}

In the planar limit the number of Feynman diagrams grows only exponentially with
order~\cite{KNN}. At finite IR and UV cutoffs, one would therefore expect many observables
to be given by a series in the 't Hooft coupling constant with a finite radius
of convergence.  We assume that this remains true in the continuum and
thermodynamic limit. 
For asymptotically free theories that would correspond to 
an expansion in scale~\cite{thooft-conv}.
This has not been proven for pure $SU(N)$ gauge theories,
the topic of this paper. In both three and four Euclidean dimensions large
distance phenomena are so different from short distance ones that one would
guess that there are points of nonanalyticity in scale determining the radius
of convergence of planar perturbation theory~\cite{neuberg-inst}. If the theory has no qualitatively 
strong scale dependence, the radius of convergence might be governed by a 
singularity corresponding to some complex valued scale, an unphysical point.
Our guess stems from the intuition that free-field short distance and long-distance
confinement are qualitatively different to such an extent, that once we work
in an approximation that produces a finite radius of convergence at short distances,
a phase transition as a function of scale is plausible. 
The simplest example would be a circular Wilson loop, expressed as a trace of a
closed loop parallel transport in the fundamental representation, as 
a function of $R$ the radius. For reasons we do not wish to be side-tracked 
into describing, it is
unlikely that a critical radius exists in this case. However, if one looks
at the distribution of the eigenvalues of the random unitary
matrix whose trace is the Wilson loop a critical radius can be identified~\cite{largen-wloop}.

Before one speaks about critical radii one needs to define renormalized 
eigenvalues, and that is somewhat complicated and involves the choice of
order $N/2$ arbitrary renormalization constants. A way to avoid the $N$
dependence of the number of renormalization constants is to look at surfaces
instead of loops and define an appropriate generalization of the Wilson loop 
operator~\cite{prev}. One needs two (likely only one) new renormalization constant
in 4D, and none in 3D.

In this paper we study the 3D case and show that the new observable 
indeed has an unambiguously defined critical size associated with it. We then briefly look at the 4D case 
to identify in detail how it differs from the 3D case. Further work
on 4D is relegated to the future.

\section{Setup}
 
We embed in $R^3$ an 
infinite cylinder $\Sigma$, defined by $x_\mu(\sigma), \mu=1,2,3$ ($\sigma$ 
is short for $\sigma_\alpha$) with $\alpha=1,2$.
\be
x_1(\sigma)=\frac{R}{2\pi}\cos\frac{2\pi \sigma_1}{l_1};
~~x_2(\sigma)=\frac{R}{2\pi}\sin\frac{2\pi \sigma_1}{l_1};
~~x_3(\sigma)=\sigma_2~~;-\pi\le\sigma_1<\pi,-\infty < \sigma_2<\infty
\ee

We define a two component gauge potential $\vpot_\alpha$ on the cylinder by
\be
\vpot_\alpha=A_\mu (x(\sigma))
\frac{\partial x_\mu}{\partial \sigma_\alpha}
\ee
The Dirac matrices on the surface are
\be
\gamma_1=\pmatrix{0 & 1\cr1 & 0}~~\gamma_2=\pmatrix{0 & -i\cr i & 0}
\ee

The new observable 
is~\footnote{$2+\epsilon, \epsilon >0$ dimensional fermions coupled to
four dimensional gauge fields were recently considered in a different
context as an example of a theory with a conformal
phase~\cite{Kaplan:2009kr}. See~\cite{Rey:2008zz}
for related earlier work.
}:
\be
Q(m,\Sigma)=\frac{
\int [dA_\mu ] [d\bar\psi d\psi ]  
e^{-\frac{1}{4g^2}\int d^3 x \tr F^2 +\int_\Sigma d^2\sigma
\bar\psi(\sigma)[\gamma_\alpha \partial_{\sigma_\alpha} - m - i\gamma_\alpha\vpot_\alpha(\sigma)]\psi(\sigma)}}{
\int [dA_\mu ] e^{-\frac{1}{4g^2}\int d^3 x \tr F^2}}
\ee
The two dimensional massless Dirac operator is:
\be
D_2 (\Sigma )=\gamma_\alpha \partial_{\sigma_\alpha} -  i\gamma_\alpha\vpot_\alpha(\sigma)
\label{diracb}
\ee

Denoting the Hermitian generators of $su(N)$ in the fundamental representation by
$T^j$, $j=1,..,N-1$ the nonabelian fermion current coupled to $\vpot$ is given by
\be
J^j_\alpha (\sigma ) = \bar\psi (\sigma) \gamma_\alpha T^j \psi (\sigma)
\ee
The abelian vector current is
\be
J_\alpha = \bar\psi (\sigma) \gamma_\alpha \psi (\sigma)
\ee
Power counting and symmetries allow for two local four-fermion counter-terms at $m=0$:
\be
\LL_1 = J^j_\alpha J^j_\alpha,~~~\LL_2= J_\alpha J_\alpha
\label{fourfermi}
\ee
Both terms can be simplified by a Fierz transformation.  
Both counter-terms are dimensionless, so there are no ultraviolet divergences worse than logarithmic. However, they won't be generated by
the three dimensional gauge interaction.

To identify possible ultraviolet divergences it is enough to replace the
closed cylinder  $\Sigma$ by
an infinite two dimensional plane. The $A_\mu$ propagator in Feynman gauge in Fourier
space induces a two dimensional propagator:
\be
\int_{p_\|^2 + p_\perp^2 \le \Lambda^2} \frac{d^2 p_\| dp_\perp}{(2\pi)^3} \frac{f(p_\|^2)}{p_\|^2 + p_\perp ^2 }=
\int_{p_\|^2 \le \Lambda^2} \frac{d^2 p_\|}{4\pi^3} f(p_\|^2) \frac{1}{\sqrt{p_\|^2}}
\tan^{-1}\sqrt{\frac{\Lambda^2}{p_\|^2}-1}
\label{nodiv}
\ee
Taking $f$ to vanish sufficiently fast at infinity, the limit $\Lambda\to\infty$ leaves a finite expression, 
\be
\int \frac{d^2 p_\|}{8\pi^2} f(p_\|^2) \frac{1}{\sqrt{p_\|^2}}
\label{sharpcutoff}
\ee
unlike in 4D, where one is left with a logarithm of $\Lambda$. 

In short, because $g^2$ has dimensions of mass this mixed 2D/3D 
theory still is super-renormalizable. The main claim is that there exists a
pure number, $r_c$, such that at $N=\infty$ for $Rg^2 N > r_c$ we have
$\frac{1}{N}\langle\bar\psi\psi\rangle\ne 0$ while for $Rg^2 N < r_c$, 
$\frac{1}{N}\langle\bar\psi\psi\rangle = 0$. At finite $N$, 
$\frac{1}{N}\langle\bar\psi\psi\rangle = 0$ for all $R$. That 
$\frac{1}{N}\langle\bar\psi\psi\rangle\ne 0$ at infinite $N$ means that
the infinite $N$ normalized single eigenvalue density of the random operator
$i D_2(\Sigma)$, $\rho(\mu)$, is nonzero at the origin: $\rho(0) >0$. 
By clustering arguments the monomials $\langle[\frac{1}{N}\bar\psi\psi]^k\rangle$ can also be evaluated 
in field theory and expressed in terms of multi-eigenvalue densities of
the operator $D_2(\Sigma)$. 
The upshot of these relations is that if $\rho(0) >0$, the individual 
low lying eigenvalues of $iD_2(\Sigma)$ are distributed in the same way
as the eigenvalues of a matrix $\pmatrix{0&W\cr W^\dagger&0}$ with the
probability density of the $n\times n$ matrix $W$ given by $\exp(-n\kappa
{\rm Tr} W^\dagger W)$ where $\kappa$ is simply related to $\rho(0)$ and $n$
is taken to infinity~\cite{chirmt}.

In this manner, the number $r_c$ identifies the smallest radius for which
the smallest eigenvalues of $-D_2^2(\Sigma)$ obey an explicitly known
distribution law. The relation to random matrix theory does not depend
on the chiral symmetry being non-anomalous~\cite{jacshifr}. 
It only requires that the symmetry
be obeyed by $D_2(\Sigma)$ gauge configuration by gauge configuration and
that in the large $N$ limit $\infty > \rho(0) \ne 0$. It is also immaterial that at infinite $N$ one can neglect the fermion determinant factor in the distribution of the gauge fields. But, this fact allows a more
direct interpretation of $Q(m,\Sigma)$ as a 
nonlocal operator in the pure gauge theory, similar to a Wilson loop, only
associated with a surface rather than a curve. At infinite $N$ the dependence
of $Q$ on $m$ becomes nonanalytic in $m$ at $m=0$ if $g^2 NR > r_c$.

\section{An abelian exercise}

Replacing the gauge group $SU(N)$ by $U(1)$ eliminates the gauge field
nonlinearities and there is no parameter $N$ to take to infinity. 
There are no phase transitions in the fermion world, 
but now the chiral symmetry is anomalous. As a result, for any $R$ one
expects a nonzero fermion condensate. In particular, 
we can replace the cylinder by an infinite plane. Chiral random matrix
theory would be as applicable as before. In spite of the apparent 
non-linearity in the gauge field fermion coupling, the condensate can
be computed by only carrying out Gaussian integrals. The calculation is
almost identical to one way of solving the pure 2D 
Schwinger model~\cite{ourchten}. 

The fermions now live on the $x_3=0$ plane and are not quenched. 
The condensate would be nonzero but also suffer from an infrared divergence if we quenched the fermions, just like in the Schwinger model \cite{Kiskis:2000sb,Damgaard:2005wz}. 

The partition 
function is given by:
\be
Z=\int [dA_\mu ] [d\bar\psi d\psi] e^{-\frac{1}{4e_0^2} \int d^3 x F_{\mu\nu}^2 +
\int d^2\sigma \bar\psi \gamma_\alpha ( \partial_\alpha - i A_\alpha ) \psi }
\ee
Here $\sigma_{\alpha}=x_\alpha, \alpha=1,2$ are coordinates on 
the plane. $e_0^2$ is the abelian gauge coupling. 

We  wish to calculate
\be
\langle \bar\psi\psi(\sigma) \bar\psi\psi (\sigma')\rangle
\ee
and obtain the condensate $\langle\bar\psi\psi\rangle$ from the limit
$|\sigma-\sigma'|\to\infty$ assuming clustering. This strategy allowed us to
set $m=0$ from the beginning. 

\subsection{Decoupling the fermions}

The gauge potential on the plane can be decomposed as:
\be
A_\alpha=\partial_\alpha \chi +\epsilon_{\alpha\beta} \partial_\beta \phi
\ee
We now change Grassmann integration variables as follows:
\be 
\psi^R=e^{-i\chi+\phi}\psi^R_f,~~\bar\psi^R=e^{i\chi-\phi}\bar\psi^R_f,~~
\psi^L=e^{-i\chi-\phi} \psi^L_f,~~\bar\psi^L=e^{i\chi+\phi}\bar\psi^L_f
\ee
Here $R,L$ denote the chiral components of the surface Euclidean Dirac fields.
The change of variables exposes the anomaly and eliminates the gauge-fermion
coupling. As a result,  
we can replace the fermion dependent part of the Lagrangian by
\be
S_F = -\int d^2\sigma \left [ \bar\psi_f \gamma_\alpha\partial_\alpha \psi_f + 
\frac{1}{2\pi} \phi\partial^2\phi \right ]
\ee
We now see that $\bar\psi_f,\psi_f$ 
are free and massless. The two dimensional scalar field 
$\phi$ is determined by the electromagnetic field strength $F_{1,2}$ restricted to
the $\sigma$-plane:
\be
\partial_\alpha \partial_\alpha \phi (\sigma) = F_{1,2}(\sigma,x_3=0)
\ee

The fermion bilinears we are interested in are $V,\bar V$ defined by
\be
V=\bar\psi^L \psi^R,~{\bar V} = \bar\psi^R \psi^L,  ~\bar\psi\psi=V+{\bar V}
\ee
In terms of the fields appearing in $S_F$ we have
\be
V(\sigma){\bar V}(0) = \bar\psi_f^L(\sigma)\psi_f^L(0)\psi^R_f (\sigma) {\bar\psi}^R_f(0) 
e^{2[\phi(\sigma) -\phi (0)]}
\label{obs}
\ee
To compute the expectation value we need the portion of the $\phi$ dependent action
coming from the pure gauge part. 

\subsection{The gauge action for $\phi$}

If $S_F=0$ the expectation value of $\phi(\sigma)\phi(0)$ is determined by the two
point function of the field strength. The gauge action contribution to the $\phi$
action is Gaussian with a kernel given by inverting the two dimensional 
kernel $\langle F_{12}(x)F_{12}(0)\rangle$ evaluated in pure gauge theory 
after setting $x_3=0$ and $x_\alpha=\sigma_\alpha $. Using $F_{\mu\nu}=\partial_\mu A_\nu - \partial_\nu A_\mu$
we have the gauge invariant expression:
\be
\langle F_{12}(\sigma)F_{12}(0)\rangle = -\frac{e_0^2}{4\pi} \frac{1}{|\sigma|^3} + 
\frac{\delta^2(\sigma)}{\sigma}{\rm ~type~of~terms}
\ee

To perform the inversion we need the two dimensional Fourier transform 
of $g(\sigma)\equiv 1/|\sigma|^3$ and require
ultraviolet regularization. We choose to regulate by:
\be
\langle F_{12}(\sigma)F_{12}(0)\rangle = -\frac{e_0^2}{4\pi} \frac{1}{(|\sigma|^2+a^2)^{\frac{3}{2}}}
+\frac{e_0^2 }{2a} \delta^2 (\sigma)
\ee
Here $a$ is a short distance cutoff, equivalent to $1/\Lambda$ where 
$\Lambda$ was the high energy cutoff used earlier. The $\delta$-function 
term is chosen to make the
integral over $\sigma$ zero, because we expect to get a contribution to 
the expectation value of~(\ref{obs}) only from gauge fields $F_{12}(\sigma)$ which integrate to zero over the plane. 
\bea
{\tilde g}(k)&=& \int d^2 \sigma \frac{1}{(|\sigma|^2+a^2)^{\frac{3}{2}}} e^{i{\vec k} \cdot {\vec \sigma}}= 2\pi \int_0^\infty
\frac{\sigma d\sigma}{[\sigma^2 + a^2]^{\frac{3}{2}}} J_0(k\sigma)\cr
&=&-\frac{2\pi}{a}\frac{d}{da}\int_0^\infty
\frac{\sigma d\sigma}{\sqrt{\sigma^2 + a^2}}J_0(k\sigma)
=\frac{2\pi}{a}e^{-ka}
\eea
Here, $k=|{\vec k}|$. We now get:
\be
\int d^2\sigma \langle F_{12}(\sigma) F_{12}(0) \rangle e^{i{\vec k}\cdot {\vec \sigma}} =
\frac{e_0^2}{2a} \left(1-e^{-ka}\right)
\label{FFthreed}
\ee 
We can take the limit of $a\to 0$ at a fixed $k$ and get
\be
\int d^2\sigma \langle F_{12}(\sigma) F_{12}(0) \rangle e^{i{\vec k}\cdot {\vec \sigma}} =
\frac{e_0^2k}{2}.
\ee
This result is compatible with~(\ref{sharpcutoff}), obtained in Feynman
gauge and using a sharp momentum cutoff $\Lambda$ at an 
intermediary stage, instead of the short distance cutoff $a$ used here.
Since
\be
F_{12}(x) = -\partial^2 \phi(x),
\ee
the gauge contribution to the $\phi$ action is
\be
S_g(\phi)=\frac{1}{e_0^2}\int d^2\sigma
\phi(x) 
\left[-\partial^2\right]^{\frac{3}{2}}
\phi(x)
\ee

\subsection{Condensate}

Combining the free fermion field propagators with the result of carrying out a Gaussian integral over $\phi$ we obtain, 
\be
\langle V(\sigma) {\bar V}(0) \rangle=\frac{1}{(2\pi)^2} \frac{1}{|\sigma|^2}
e^{4 F\left(\frac{e_0^2|\sigma|}{2\pi}\right)},
\label{keyaa}
\ee
where
\be
F\left(\frac{e_0^2|\sigma|}{2\pi}\right)=
\int \frac{d^2k}{(2\pi)^2} \frac{1-e^{ik\cdot\sigma}}{\frac{k^2}{\pi}
+\frac{2k^3}{e_0^2}}
=\frac{e_0^2|\sigma|}{4\pi} \int_0^\infty \frac{du}{u}\frac{1-J_0(u)}{\frac{e_0^2|\sigma|}{2\pi}+u}.
\ee
Defining
\be
z=\frac{e_0^2 |\sigma|}{2\pi},
\ee
we obtain
\be
2F(z) = \gamma - \ln 2 + \ln z + \int_0^\infty dt
\frac{e^{-zt}}{\sqrt{1+t^2}}
\ee

The dimensionless correlation function $G(z)$ is given by:
\be
G(z) = \frac{(2\pi)^2}{e_0^4} \langle V(\sigma) {\bar V}(0) \rangle
=\frac{1}{(2\pi z)^2} e^{4F(z)}=
\left[ \frac{e^\gamma}{4\pi} e^{\int_0^\infty dt
   \frac{e^{-zt}}{\sqrt{1+t^2}}}
\right]^2
\ee
The integral in the above equation asymptotically goes like
$\frac{1}{z}$
and therefore 
\be
\lim_{z\to\infty} G(z) = \left[ \frac{e^\gamma}{4\pi}\right]^2
\ee
and we have a nonzero, finite, condensate.
At short distances, the integral goes like $-\ln z$ and therefore the propagator goes
like $\frac{1}{z^2}$, the expected short distance behavior. 

No renormalization was needed and we got a finite condensate indicating
that the spectrum of the two dimensional Dirac operator on the surface
will exhibit chiral random matrix behavior as a result of its dependence
on the background abelian gauge field. That gauge field is inherited from
the 3D bulk, but its fluctuations are augmented by the feedback from the
two dimensional fermions, via the anomaly term. 

\section{Condensate in the non-abelian case}

We wish to show that one will get a nonzero finite condensate as 
above also in the $SU(N)$ case and in the infinite $N$ limit, so without
the feedback of the fermion determinant. This can be done only numerically.

We work on cubic lattices of $L^3$ sites with periodic boundary conditions 
on which live $SU(N)$ gauge variables. The action is the standard single 
plaquette Wilson action and the lattice 't Hooft coupling is denoted by $b$.
Large $N$ reduction says that the infinite $N$ limit for any $L>L_c(b)$ becomes
$L$ independent~\cite{reduction}. The $L,b$ pairs at which we carried out simulations satisfy
$L>L_c(b)$. 

We first put fermions on one of the $1-2$ planes with antiperiodic 
boundary conditions. The overlap action couples them to the gauge link variables
in the particular plane~\cite{overlap}. For each statistically independent gauge configurations, determined by the three
dimensional Wilson action we calculate the two lowest positive eigenvalues 
$\lambda_1<\lambda_2$ 
of the overlap operator. We ascertain that there is chiral random matrix
behavior and extract the condensate. This is repeated for several choices of $b$.
We then check whether the condensate scales with $b$ in the way expected of a
physical quantity of dimension mass.  We find that the answer is positive.

The eigenvalues are computed using the Ritz algorithm and are 
used to estimate the condensate by formulas from chiral random 
matrix theory \cite{Damgaard:2000ah,Narayanan:2004cp}:
\be
\Sigma_1 = \frac{1.722}{2 m_{\rm tad}\langle\lambda_1\rangle NV},\ \ 
\Sigma_2 = \frac{4.791}{2 m_{\rm tad}\langle\lambda_2\rangle NV},\ \ 
m_{\rm tad} = \frac{m_w-2(1-u)}{u},\ \  
u^4\equiv e=\frac{1}{N}\langle \Tr U_p\rangle,
\ee
where $U_p$ is the parallel transporter around a plaquette and $m_w$
is the Wilson mass parameter appearing in the overlap Dirac operator and
$V=L^2$.
Chiral random matrix theory is diagnosed by the two determinations
$\Sigma_j$ agreeing with each other. 

Our results for the condensate are presented in Table~\ref{tab4}.
For $b=0.60$ we have shown evidence for reduction since the results
on $3^3$ agree with $4^3$. 
Another example of reduction 
can be seen at $b=0.90$ by comparing results on $5^3$ with $6^3$. At $b=0.80$, we obtain consistent estimates
from $N=59$ on $4^3$ and $N=47$ on $5^3$ showing that we are
in the large $N$ and large volume limit.

\TABLE{
\begin{tabular}{ccccccc}
\hline
$L$ & $N$ & $b$ & $e$ &
$\frac{1}{0.37674}\langle\frac{\lambda_1}{\lambda_2}\rangle$ &
$\Sigma_1$ & $\Sigma_2$\\
\hline
$3$ & $47$ & $0.45$ & $0.52558(17)$ & $0.983(19)$ & $0.1869(40)$ & $0.1836(18)$\\
$3$ & $47$ & $0.50$ & $0.59914(11)$ & $1.020(18)$ & $0.1489(29)$ & $0.1527(14)$\\
$3$ & $47$ & $0.55$ & $0.64732(10)$ & $0.995(20)$ & $0.1326(29)$ & $0.1320(13)$\\
$3$ & $47$ & $0.60$ & $0.68372( 8)$ & $1.018(20)$ & $0.1229(28)$ & $0.1264(14)$\\
$4$ & $47$ & $0.60$ & $0.68355( 5)$ & $0.997(17)$ & $0.1196(22)$ & $0.1197(10)$\\
$4$ & $47$ & $0.70$ & $0.73632( 4)$ & $1.012(17)$ & $0.1001(18)$ & $0.1014( 8)$\\
$4$ & $59$ & $0.80$ & $0.77335( 3)$ & $1.009(16)$ & $0.0842(16)$ & $0.0858( 8)$\\
$5$ & $47$ & $0.80$ & $0.77338( 3)$ & $1.006(15)$ & $0.0864(13)$ & $0.0863( 6)$\\
$5$ & $47$ & $0.90$ & $0.80115( 2)$ & $1.020(15)$ & $0.0738(12)$ & $0.0756( 6)$\\
$6$ & $47$ & $0.90$ & $0.80106( 2)$ & $1.007(14)$ & $0.0758(11)$ & $0.0764( 5)$\\
$6$ & $47$ & $1.00$ & $0.82256( 1)$ & $1.018(14)$ & $0.0653(10)$ & $0.0665( 4)$\\
$6$ & $47$ & $1.10$ & $0.83987( 1)$ & $1.020(14)$ & $0.0576( 9)$ & $0.0595( 4)$\\
$8$ & $47$ & $1.20$ & $0.85402( 1)$ & $1.024(13)$ & $0.0529( 7)$ & $0.0546( 3)$\\
$8$ & $47$ & $1.30$ & $0.86587( 0)$ & $1.036(12)$ & $0.0480( 6)$ & $0.0498( 3)$\\
$8$ & $47$ & $1.40$ & $0.87594( 0)$ & $1.024(12)$ & $0.0449( 6)$ & $0.0461( 3)$\\
$9$ & $47$ & $1.50$ & $0.88458( 0)$ & $0.998(11)$ & $0.0437( 5)$ & $0.0435( 2)$\\
\hline
\end{tabular}

\caption{\label{tab4} Condensate for a plane, at
several values of $b$ and $N$ 
on an $L^3$ lattice.}
}
\TABLE{
\begin{tabular}{cccc}
\hline
$N$ & $\langle \frac{\lambda_1}{\lambda_2} \rangle$ & $\Sigma_1$ &
$\Sigma_2$ \\
\hline
13 & 0.423(6) & 0.0689(10) & 0.0787(5) \\ 
23 & 0.402(5) & 0.0681(10) & 0.0735(5) \\
29 & 0.401(5) & 0.0672(10) & 0.0726(5) \\
37 & 0.389(5) & 0.0684(10) & 0.0712(5) \\
47 & 0.390(5) & 0.0686(10) & 0.0713(5) \\
101 & 0.378(5) & 0.0681(10) & 0.0683(5) \\
\hline
\end{tabular}
\caption{\label{tab5} Results showing the 
approach to chiral random matrix theory 
for a cylinder with a $4\times 4$ square base embedded 
on a $5^4$ lattice at
$b=0.7$.}}

Since the three dimensional
coupling, $b$, has dimensions of length, we expect the condensate
to behave as $\frac{1}{b}$ as one goes to the continuum limit of
the three dimensional gauge theory if the 2D scale is set by the three
dimensional bulk scale.
The condensate is plotted as a function of the tadpole improved
coupling in Fig.~\ref{fig5}. There is a curvature in the behavior
indicating a sizable finite lattice spacing effect. 
To extrapolate to the 
continuum limit we used the so called tadpole improved 
coupling $b_I$ instead of $b$ ($b_I=e b$). 
A fit to the sum of two terms
proportional to
$b_I^{-1}$ and $b_I^{-2}$ respectively agrees well with the data. 
\vskip 1.5cm

\FIGURE{
\centering
\label{fig5}
\centering
\includegraphics[scale=0.5]{cond23d.eps}
\caption{ The data for the condensate is in Table~\ref{tab4}. The
  $b=0.6$ result on $3^3$ lattice at $N=47$, the $b=0.8$ result on
$4^3$ and $N=59$ and the $b=0.9$ result on $5^3$ and $N=47$
have not been included in the plot.
The fit includes a finite lattice spacing
correction of the form $b_I^{-2}$.}}

Two quarks on the plane will feel a linear potential at large distances 
reflecting confinement in 3D $SU(N)$ YM theory. We know what the condensate
should be if this were a 2D $SU(N)$ YM theory with the same string tension.
When we compare the measured condensate to that of the hypothetical 2D theory
we find that the condensate $\Sigma$ in units of the string tension $\sigma$,
that is the ratio $\Sigma/\sqrt{\sigma}$, is 0.29 in the present case, whereas
the exact value in purely 2D gauge theory is 0.23. In Casher's picture of
spontaneous chiral symmetry breaking the condensate is determined by the size
of a would-be massless bound 
state of two massless quarks in an $s$-wave~\cite{casher}. A smaller
bound state corresponds to a larger condensate. Within this picture we would then say
that the size of these would-be bound state is smaller when the gauge forces
have a 3D origin than when they have a 2D origin. In the 3D case the would-be 
bound state forms at a scale shorter than the scale at which confinement alone would have formed it. We can say this because in the 2D case the forces are purely confining.
The comparison is possible because the kinematics of the fermions are identical
in the two case. Since the 3D forces at shorter distances are strongly attractive
the fact that $\Sigma/\sqrt{\sigma}$ is larger for 3D forces than for 2D forces makes
sense. Actually, in this context, had we found the opposite relation, namely  $\left ( \frac{\Sigma}{\sqrt{\sigma}}\right )_{3d} <
\left (\frac{\Sigma}{\sqrt{\sigma}}\right )_{2d}$, we could have claimed to have
invalidated Casher's argument. Conversely, one could interpret  
the relative 
closeness of $0.29$ to $0.23$ as evidence that confinement makes a major
contribution to the condensate in our 2D/3D model. 

\section{The large $N$ phase transition for the cylinder}

The lattice version of the cylinder has an $s\times s$ square as
basis in the $1-2$ plane. The fermions live on the surface of this tower and
interact with the gauge fields on the links in the surface by the chirally invariant
overlap action. The fermions obey antiperiodic boundary conditions round the 
$s\times s$ loops and in the $3$ direction.

For each statistically independent gauge configuration, determined by the three
dimensional Wilson action we calculate the 
two lowest eigenvalues of the square
of the overlap operator, $0<\lambda_1^2<\lambda_2^2$, for the $L^2$ cylinders in one selected direction we call
$3$. The collected data produces histograms which are used to determine whether the
distribution of the eigenvalues for given values of $s,b$ becomes that of chiral
random matrix theory in the large $N$ limit, or, instead, indicates a positive
spectral gap. In the former case, we extract a lattice value of the condensate as
before. For a given $s$ we shall find non-zero condensates so long
as $b<b_c(s)$. To find $b_c(s)$ we fit the condensate to $K\sqrt{b_c(s)  - b}$
in a regime of $b$-values which is close, but not too close to $b_c(s)$. We then
check whether $b_c(s)$ scales as expected with $s$ and find that it does, 
eventually producing
a physical length for the critical length $s$ of the base square of the cylinder.
We looked at square bases of lengths $s=3,4,5$.

\subsection {$4\times 4$ square base -- Details}

Table~\ref{tab5} displays the approach to chiral random matrix theory at
$b=0.7$. The average of the eigenvalue 
ratio, $\langle\frac{\lambda_1}{\lambda_2}\rangle$, 
approaches the chiral random matrix theory
prediction slowly and attains it only at $N=101$. 
The estimate of the condensate obtained from the 
first eigenvalue 
changes little over the entire 
range of $N$ values listed in Table~\ref{tab5} but the second eigenvalue
takes much longer to converge.
A plot of the distribution of the ratio of the eigenvalues in
Fig.~\ref{fig6} exhibits a slow convergence to that predicted by chiral random matrix theory.

\FIGURE{
\centering
\includegraphics[scale=0.5]{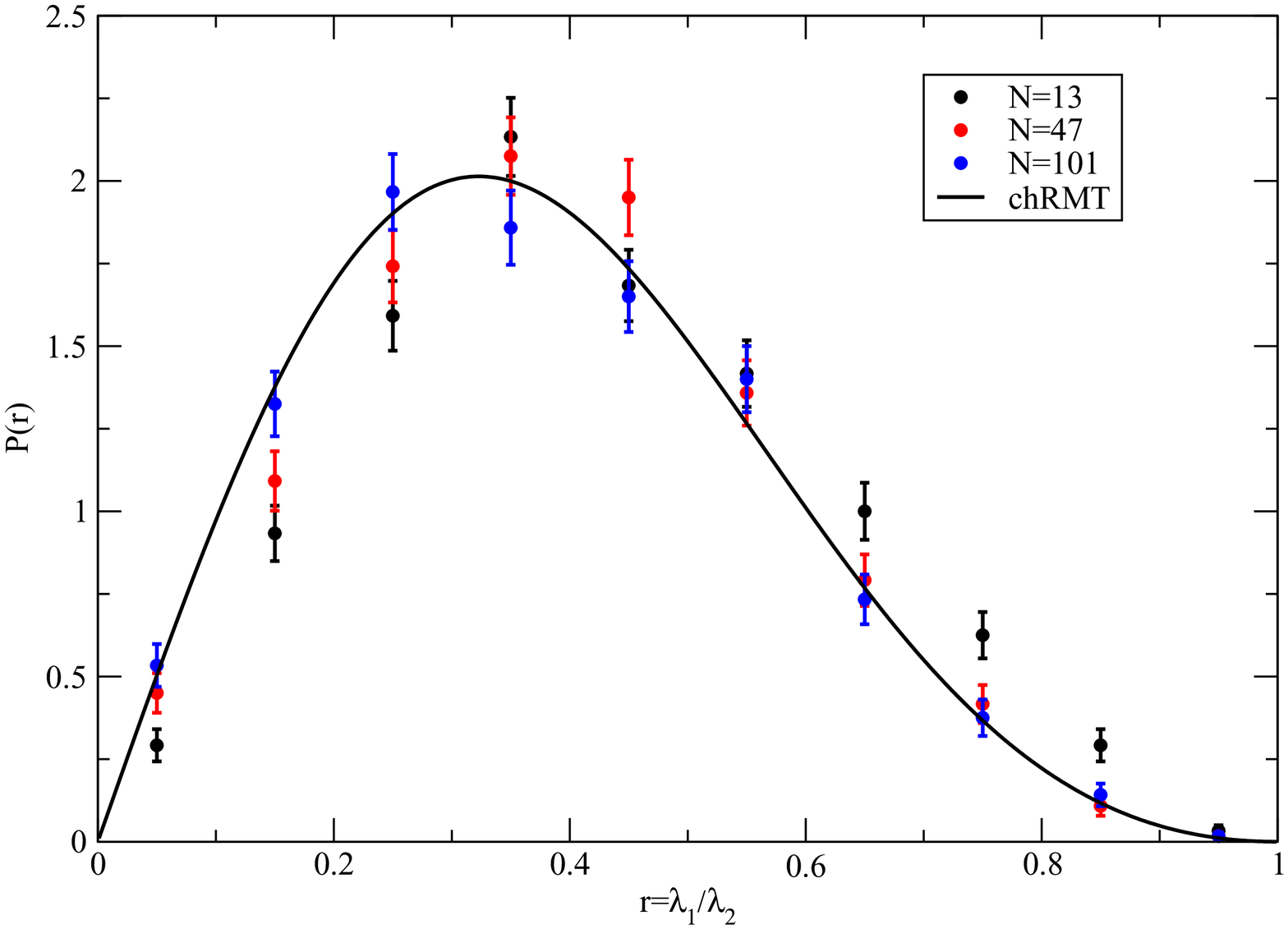}
\caption{\label{fig6}Distribution of $\frac{\lambda_1}{\lambda_2}$ 
at $b=0.7$ and $N=13,47,101$ for a 
cylinder with a $4\times 4$ square base 
embedded in a $5^3$ lattice.}
}

We computed the condensate for several values of the coupling below
$b=0.7$
in order to study the approach to the critical point. Our estimates
are listed in Table~\ref{tab6}. 

\TABLE{
\begin{tabular}{cccccc}
\hline
$b$ & $N$ & $L$ & $\frac{1}{0.37674}\langle\frac{\lambda_1}{\lambda_2}\rangle$ &
$\Sigma_1$ &
$\Sigma_2$ \\
\hline
0.60 & 47 & 4 & 1.02(2) & 0.102(2) & 0.1030(9) \\ 
0.62 & 79 & 5 & 0.99(1) & 0.099(2) & 0.0981(7) \\
0.64 & 79 & 5 & 0.98(2) & 0.093(2) & 0.0919(9) \\
0.66 & 79 & 5 & 1.01(2) & 0.083(2) & 0.0836(8)\\
0.68 & 79 & 5 & 1.01(2) & 0.074(2) & 0.0751(7) \\
0.70 & 101 & 5 & 1.00(1) & 0.068(1) & 0.0683(5) \\
\hline
\end{tabular}
\caption{\label{tab6} Results showing the estimates for the condensate
for a cylinder with a $4\times 4$ square base embedded
in a $L^3$ lattice and at various $N$.}
} 

All results are consistent with chiral
random matrix theory as the condensates predicted by the two eigenvalues agree 
with each other. A plot of the square of the condensate
versus the bare coupling (we could have also used
the tadpole improved coupling) in Fig.~\ref{fig7} 
shows that the behavior is linear indicating
that the critical exponent is $\frac{1}{2}$. The critical value of
the bare 
coupling for a $4\times 4$ square base is estimated at 0.77(3).

\FIGURE{
\centering
\includegraphics[scale=0.5]{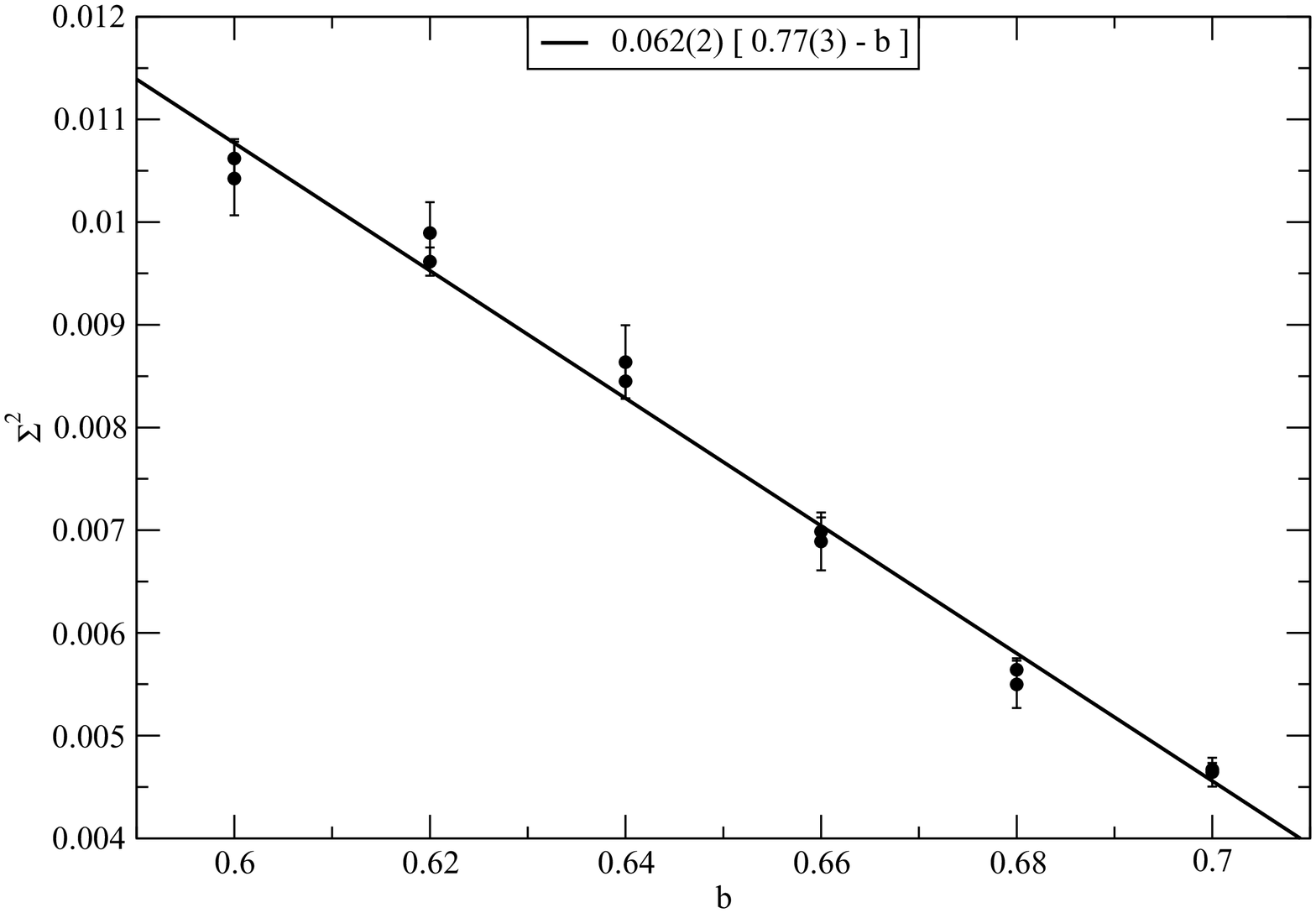}
\caption{\label{fig7}A plot of the condensate listed in Table~\ref{tab6} as a
function of the bare coupling.}
}

We studied the behavior of the two lowest eigenvalues at $b=0.80$ for
several values of $N$ on a $5^3$ lattice in order to see if it is in
the
symmetric phase. A plot of the data is shown in Fig.~\ref{fig8}. 
The gap at infinite $N$ is estimated by a fit to a $N^{-\frac{2}{3}}$ 
(a standard soft edge random hermitian matrix prediction) 
plus a subleading $N^{-1}$ correction which  reproduces the data quite well
as shown in Fig.~\ref{fig8}. The fit forces the $N=\infty$ limits of the two
eigenvalues to be identical. 
The estimated gap at infinite $N$ is small, 
0.00167(4), but still 
indicates that $b=0.80$ is in the symmetric phase. The subleading term
is essential for the fit to work and this indicates that $b=0.80$ is close
to the transition point.

\FIGURE{
\centering
\includegraphics[scale=0.5]{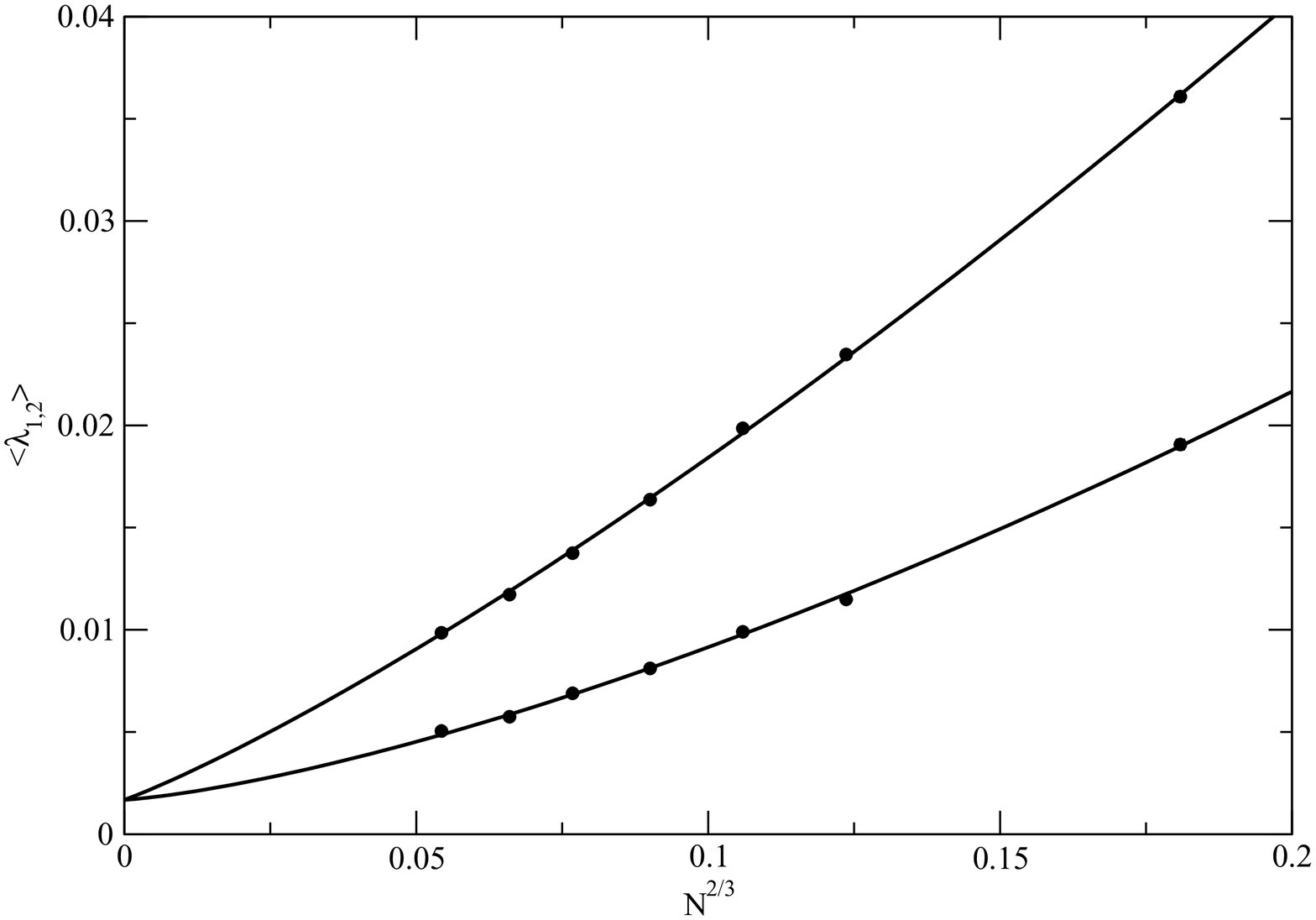}
\caption{\label{fig8}A combined fit of $\langle\lambda_{1,2}\rangle$ 
at $b=0.8$ and $N=13,23,29,37,47,59,79$ for a 
cylinder with a $4\times 4$ square base 
embedded in a $5^3$ lattice.}
}

We now wish to compare the large $N$ character of this transition to that we
have studied for Wilson loops in the past. The essence of the latter was that 
close to the transition the
eigenvalue separation in the least populated regime went as $\frac{1}{N}$ for large 
scales and as $\frac{1}{N^{2/3}}$ for
small scales, while, exactly at the critical size, it went 
as $\frac{1}{N^{3/4}}$. 
Something similar happens to the level separation in our case:
In figure~\ref{fig20} we show fits of the logarithm of the eigenvalue difference to 
a linear function of $\log(N)$ with fitted slope and intercept. The fits 
cease being stable when subleading terms are added as the ranges of $\log(\lambda_2-\lambda_1)$
and $\log(N)$ are not large enough. The two extreme couplings, $b=0.70$ and $b=0.80$ are close 
to the transition from different sides. Farther from the transition, in particular
for smaller $b$, the slope in the fit would match better the appropriate random
matrix expectation. As $b$ goes through the transition the effective slope one gets from
any fit to a set of data taken at finite $N$, would have to vary smoothly from -1 to -2/3.
The change would be steepest at the critical $b$-value. One could define and effective
$b_c(N)$ where $N$ is the largest value in a well defined set used to get the linear fit and where
the effective slope is exactly -3/4.  

However, one should keep in mind that in the case of Wilson loops we
dealt with an odd dimensional Dirac operator (living on the loop), while
here we deal with an even dimensional Dirac operator. We expect 
that the correct random matrix models describing the low 
eigenvalues of the Dirac operator in the ungapped phase would differ
in the two cases because chirality exists only in even dimensions. 
For odd dimensions, see~\cite{jacisma}.

\FIGURE{
\centering
\includegraphics[scale=0.5]{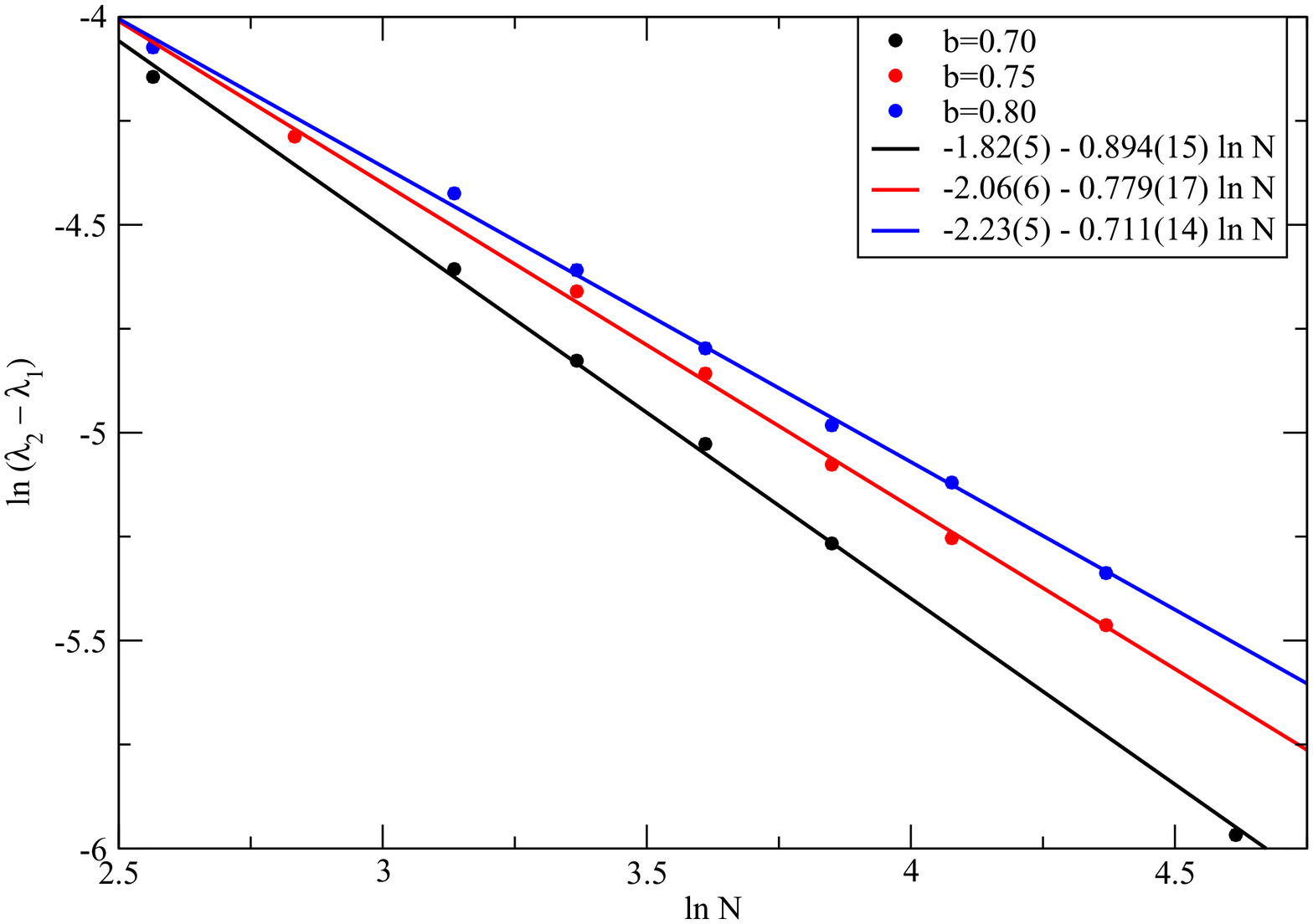}
\caption{\label{fig20} Logarithm of lowest positive eigenvalues difference
versus $\log(N)$ and linear fits. The data is at couplings $b=0.70,0.75,0.80$
for a cylinder with a $4\times 4$ square base embedded in a $5^3$ lattice.}
}


\subsection{Evidence for a continuum critical size}

In order to check whether the critical size also 
has a continuum limit, we ran simulations of cylinders 
with $3\times 3$ and $5\times 5$ bases. 
Our presentation of 
the $3\times 3$ and
$5\times 5$ cases will be less detailed than that of the $4\times 4$ case. 

We set $N=47$ and used $4^3$ lattices for
the cylinders having a $3\times 3$ base. 
Cylinders with a $5\times 5$ base were embedded in a $5^3$ lattice
with $N=59$,
as the lattice spacing at the transition is smaller now.  
We computed the condensate for several couplings and the
results are listed in Table~\ref{tab8}.
All the data in this table
are consistent with chiral random matrix theory.

\TABLE{
\begin{tabular}{ccccc}
\hline
$s\times s$& 
$b$ & $\frac{1}{0.37674}\langle\frac{\lambda_1}{\lambda_2}\rangle$ &
$\Sigma_1$ &
$\Sigma_2$ \\
\hline
3$\times$3 &0.45 & 1.00(1) & 0.177(3) & 0.176(1) \\ 
3$\times$3 &0.46 & 0.98(2) & 0.168(3) & 0.166(1) \\
3$\times$3 &0.48 & 0.99(2) & 0.154(3) & 0.154(1) \\
3$\times$3 &0.50 & 1.00(2) & 0.135(3) & 0.135(2)\\
3$\times$3 &0.52 & 1.00(1) & 0.125(2) & 0.125(1) \\
3$\times$3 &0.54 & 0.97(2) & 0.114(2) & 0.111(1) \\
3$\times$3 &0.55 & 1.01(2) & 0.105(3) & 0.106(1) \\
5$\times$5 &0.67 & 0.99(1) & 0.0954(14) & 0.0943(7) \\ 
5$\times$5 &0.70 & 1.00(1) & 0.0881(13) & 0.0884(6) \\
5$\times$5 &0.72 & 0.99(1) & 0.0822(12) & 0.0817(6) \\
5$\times$5 &0.75 & 1.01(1) & 0.0785(12) & 0.0793(6)\\
5$\times$5 &0.78 & 1.01(1) & 0.0680(10) & 0.0689(5) \\
5$\times$5 &0.80 & 1.00(2) & 0.0650(10) & 0.0652(5) \\
\hline
\end{tabular}
\caption{\label{tab8} Results showing the estimates for the condensate
for cylinders with $3\times 3$ and $5\times 5$ square bases embedded
in $4^3$ and $5^3$ lattices with $N=47$ and $N=59$ respectively.}
}

In order to see evidence for a continuum critical size we plot the
square of the dimensionless condensate, $\Sigma s$, as a function of
the dimensionless inverse size of the loop, $b_I/s$, for an $s\times
s$ loop in Fig.~\ref{fig9} for $s=3,4,5$. The plot for $s=4$ is the
same as in Fig.~\ref{fig7} in terms of the new variables. We see
reasonable evidence for scaling and our combined estimate for
the critical size is $s_c \sqrt{\sigma}= 1.37(10)$; for the string
tension $\sigma$, we have used the approximate formula 
$\sqrt{\sigma} b_I = \sqrt{\frac{1}{8\pi}}$~\cite{Kiskis:2008ah}.

\FIGURE{
\centering
\includegraphics[scale=0.5]{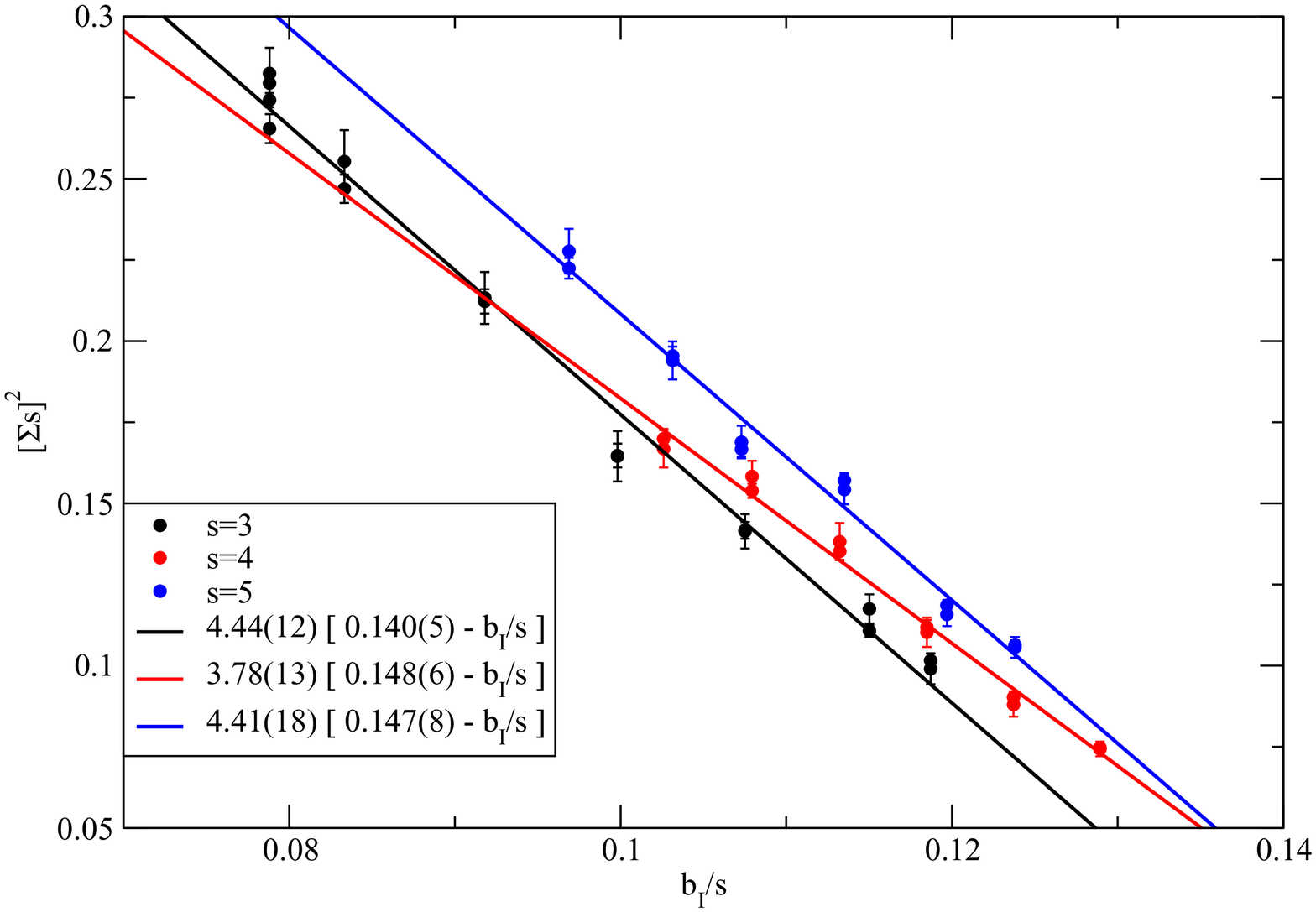}
\caption{\label{fig9}A plot of the square of the
dimensionless condensate listed in Table~\ref{tab6} and Table~\ref{tab8} as a
function of the tadpole improved coupling.}
}

\section{Four dimensions}

For Wilson loops we know that the description of the large $N$ transition 
is the same in 2,3,4 dimensions. We have seen that the large $N$ 
critical properties of the Wilson loop 
transition also extend to our new surface observable in 3 dimensions. 
So, we go to four dimensions, where renormalization is less trivial. 

\subsection{Abelian exercise}

As before, we start with the an abelian exercise. The method of solution is 
identical to that used in the three dimensional case, 
so a brief description of the differences suffices. The difference
comes in through the photon propagator restricted to the plane. Equation~(\ref{FFthreed}) is now replaced by
\be
\int d^2\sigma \langle F_{12}(\sigma) F_{12}(0) \rangle e^{i{\vec k}\cdot {\vec \sigma}} =
-\frac{e_0^2 k^2}{\pi} \left [ \frac{K_1(ka)}{ka} -\frac{1}{(ka)^2}\right ] \sim -\frac{e_0^2}{4\pi} k^2 \left [ \log \left ( \frac{ka}{2}\right )^2 +{\cal O}(ka)\right ]
\label{FFfourd}
\ee
The modified Bessel function $K_1(z)$, for $z\ge 0$ is positive and monotonically decreasing.
For $z\to 0^+$ we have
\be
K_1(z) = (z/2) \log (z/2)  +1/z +{\cal O}(z)
\ee
For $z\to\infty$ we have
\be
K_1(z) = \sqrt{\frac{\pi}{2z}}e^{-z}[1+{\cal O}(z^{-1})]
\label{highmom}
\ee
The cutoff $a$ cannot be eliminated. 
Keeping only the leading term in~(\ref{FFfourd})
we get the same expression we would have gotten had we used a sharp momentum
cutoff in Feynman gauge, $\Lambda=2a^{-1}$:
\be
\int_{p_\|^2 + p_\perp^2 \le \Lambda^2} \frac{d^2 p_\| d^2p_\perp}{(2\pi)^4} \frac{f(p_\|^2)}{p_\|^2 + p_\perp ^2 }=
\int_{p_\|^2 \le \Lambda^2} \frac{d^2 p_\|}{(2\pi)^3} f(p_\|^2) \log[\Lambda/|p_\||]
\label{logdiv}
\ee
This impacts the correlation of the bilinears:
\be
\langle V(\sigma) {\bar V}(0) \rangle=\frac{1}{(2\pi)^2} \frac{1}{|\sigma|^2}
e^{4 F(|\sigma|)}
\label{keyab}
\ee
with
\be
F(|\sigma|)=\frac{1}{2} \int_0^s \frac{du}{u} \;\; \frac{1-J_0 (u)}{1-\frac{p}{\log(u/s)}};~~~~
p\equiv 2\pi^2/ e_0^2;~~~~~~s\equiv \Lambda|\sigma|
\ee
We are interested in the large $s$ behavior of $F$, now viewed as a function of $s$.

To this end we define:
\be
\Delta F(s,p) = F(s)|_{p=0} - F(s)|_p=\frac{p}{2} \int_0^1 \frac{dx}{x} [1-J_0 (sx)]
\frac{1}{p-\log (x)}
\ee
Up to terms vanishing as $s\to\infty$ one has
\be
F(s)|_{p=0}=  \frac{1}{2}[\log(s/2) +\gamma]
\ee
and
\be
\Delta F(s,p) = \frac{p}{2}\;\log\left [ \frac{1}{p}\;\log(s)\right ]
\label{finform}
\ee

\subsection{Condensate on the plane}
We therefore find that
\be
\Lambda^{-2}\langle V(\sigma) {\bar V}(0) \rangle \sim \left [ \; \frac{e^\gamma}{4\pi}\;\left ( \frac{e_0^2}{2\pi^2} \; \log (\Lambda |\sigma| ) \right )^{-\frac{2\pi^2}{e_0^2}}\right ]^2
+{\rm~terms~that~vanish~as~}\Lambda|\sigma|\to\infty
\label{result}
\ee
At finite $\Lambda$
the condensate vanishes. However, the decay of the correlation function of bilinears 
with distance is slow. 
The coupling $e_0$ does not renormalize far from the fermion world, where it cannot be
screened. So, we cannot just eliminate the logarithmic dependence on $\Lambda$ in~(\ref{result}) by introducing a $\Lambda$ dependence in $e_0$. 

To proceed in our search for a universal continuum limit we
may investigate the
effects of a Thirring coupling, the single four fermion
term possible in the abelian case.

\subsection{Thirring term}

A Thirring term of the form
\be
S_T=\frac{g^2}{2}j_\alpha j_\alpha,~~~j_\alpha\equiv \bar\psi \gamma_\alpha\psi,~~~\alpha=1,2
\ee
is added to the action (the exponent of the integrand in the path integral is given by
$-S_T$).

By a Hubbard-Stratonovich transformation this is implemented by shifting the 
electromagnetic potential
$A_\alpha$ coupled to the fermion current to
\be
A_\alpha\to A_\alpha + B_\alpha
\ee
The action for $A_\alpha$ is as before and that for $B_\alpha$ is
\be
S_B(B)=\frac{1}{2g^2} \int d^2\sigma B_\alpha^2
\ee
$B_\alpha$ lives only on the 2D plane.

Writing
\be
B_\alpha =\partial_\alpha \chi' +\epsilon_{\alpha\beta} \partial_\beta \phi'
\ee
we see that the scalar field entering the transformation to free fermion fields
is now $\varphi$ with
\be
\varphi= \phi + \phi'
\ee
Therefore the field entering the vertex operators ${\bar V},V$ is $\varphi$ and the
transformation of integration variables from interacting to free fermion fields makes 
a contribution to the action given by
\be
S_F =-\frac{1}{2\pi} \int \varphi \partial^2 \varphi
\ee

For $\phi$ we have the electromagnetic action we found before:
\be
S_G = \frac{2\pi}{e_0^2} \int \phi \frac{\partial^2}{\log(-\partial^2/\Lambda^2)}\phi
\ee
The Thirring interaction for $\phi'$ is
\be
S_T =-\frac{1}{2g^2} \int \phi'\partial^2\phi'
\ee

We now change the scalar field integration variables, $\phi,\phi'$ to $\varphi,\phi$ and
integrate $\phi$ out. The resulting interaction for $\varphi$ is
\be
S_\varphi (\varphi) =\frac{1}{2}\int\varphi \left [
-\frac{\partial^2}{\pi} -\frac{\partial^2}{g^2} + \frac{1}{g^2} 
\frac{\partial^2}{1-\frac{2\pi g^2}{e_0^2} \frac{1}{\frac{1}{2}\log(-\partial^2 /\Lambda^2)}}
\right ]\varphi
\ee

Tracing this through amounts to replacing $F$ in eq.~(\ref{keyab}) by
\be
F(|\sigma|) =\frac{1}{2} \int_0^s \frac{du}{u}\frac{1-J_0(u)}{1-\frac{p}{\log(u/s) - g^2 p/\pi}}
\label{fullF}
\ee
where, $s=\Lambda|\sigma|$ as before. 

For $p=0$ we get the same expression as before, the Thirring term being shielded away.
Our new $\Delta F$ is
\be
\Delta F(s,p,g^2) = F(s)|_{p=0} - F(s)|_p=\frac{p}{2} \int_0^1 \frac{dx}{x} [1-J_0 (sx)]
\frac{1}{p(1+g^2/\pi)-\log (x)}
\label{deltaF}
\ee

So long as $g^2$ is kept cutoff independent, up to terms vanishing at infinite UV cutoff, 
the answer is
\be
\Delta F(s,p)=\frac{p}{2} \log\left [ \frac{1}{p(1+g^2/\pi )} \log(s) \right ]
\ee

\subsection{No continuum limit in the four dimensional Abelian case}

We realize that both 
$e_0$ and $g^2$ do not renormalize and there is no physical infinite cutoff limit. The singular set-up of the problem in which the fermions are restricted to a zero 
thickness surface does not seem to have a universal description: the surface
must be given some thickness and then continuum physics will depend on some
of the details for how this was done. In three dimensions there was no such
problem. 
\TABLE{
\begin{tabular}{cccclc}
\hline
$N$ & $b$ & $e$ &
$\frac{1}{0.37674}\langle\frac{\lambda_1}{\lambda_2}\rangle$ &
$\ \ \ \ \ln\Sigma_1$ & $\ln\Sigma_2$\\
\hline
$13$ & $0.346$ & $0.52286(10)$ & $0.993( 9)$ & $-1.893(10)$ & $-1.897(4)$\\
$23$ & $0.346$ & $0.51513( 5)$ & $1.009( 7)$ & $-1.863( 7)$ & $-1.855(3)$\\
$29$ & $0.346$ & $0.51342( 3)$ & $0.997( 7)$ & $-1.842( 7)$ & $-1.846(3)$\\
$13$ & $0.348$ & $0.53097(10)$ & $0.996( 9)$ & $-1.934(10)$ & $-1.938(4)$\\
$23$ & $0.348$ & $0.52481( 4)$ & $1.001( 7)$ & $-1.905( 7)$ & $-1.907(3)$\\
$37$ & $0.348$ & $0.52290( 2)$ & $1.000( 7)$ & $-1.900( 7)$ & $-1.900(3)$\\
$13$ & $0.350$ & $0.53771( 7)$ & $1.012( 9)$ & $-1.985(10)$ & $-1.971(4)$\\
$23$ & $0.350$ & $0.53239( 3)$ & $1.004( 9)$ & $-1.941( 9)$ & $-1.941(4)$\\
$13$ & $0.353$ & $0.54653( 8)$ & $0.999( 9)$ & $-2.016(10)$ & $-2.015(4)$\\
$23$ & $0.353$ & $0.54183( 2)$ & $1.013( 6)$ & $-2.005( 7)$ & $-1.992(3)$\\
$13$ & $0.355$ & $0.55178( 6)$ & $1.009( 9)$ & $-2.045(10)$ & $-2.034(4)$\\
$23$ & $0.355$ & $0.54741( 3)$ & $1.005( 9)$ & $-2.031(10)$ & $-2.026(4)$\\
$13$ & $0.358$ & $0.55911( 7)$ & $0.998( 9)$ & $-2.074(10)$ & $-2.074(4)$\\
$23$ & $0.358$ & $0.55508( 2)$ & $0.999( 7)$ & $-2.060( 7)$ & $-2.062(3)$\\
$13$ & $0.363$ & $0.57010( 7)$ & $1.007( 9)$ & $-2.160(10)$ & $-2.151(4)$\\
$23$ & $0.363$ & $0.56646( 2)$ & $1.004( 7)$ & $-2.127( 7)$ & $-2.127(3)$\\
$29$ & $0.363$ & $0.56587( 1)$ & $0.994( 7)$ & $-2.121( 7)$ & $-2.128(3)$\\
$13$ & $0.367$ & $0.57810( 6)$ & $1.009( 9)$ & $-2.188( 9)$ & $-2.172(4)$\\
$23$ & $0.367$ & $0.57482( 2)$ & $1.011( 9)$ & $-2.143( 9)$ & $-2.133(4)$\\
$29$ & $0.367$ & $0.57412( 2)$ & $1.015( 6)$ & $-2.189( 7)$ & $-2.173(3)$\\
$37$ & $0.367$ & $0.57372( 1)$ & $0.996( 6)$ & $-2.179( 7)$ & $-2.181(3)$\\
\hline
\end{tabular}
\caption{\label{tab3} Condensate for a plane computed at
several values of $b$ and $N$ 
on $10^4$ lattice.}
} 

In the four dimensional non-abelian case there is 
a new four fermion coupling which will renormalize and the gauge
coupling is absorbed by dynamical scale generation in the bulk. 
We hope that these ingredients will provide for a nontrivial
continuum limit in this case at the expense of one new free parameter. 

\FIGURE{
\label{fig4}
\centering
\includegraphics[scale=0.5]{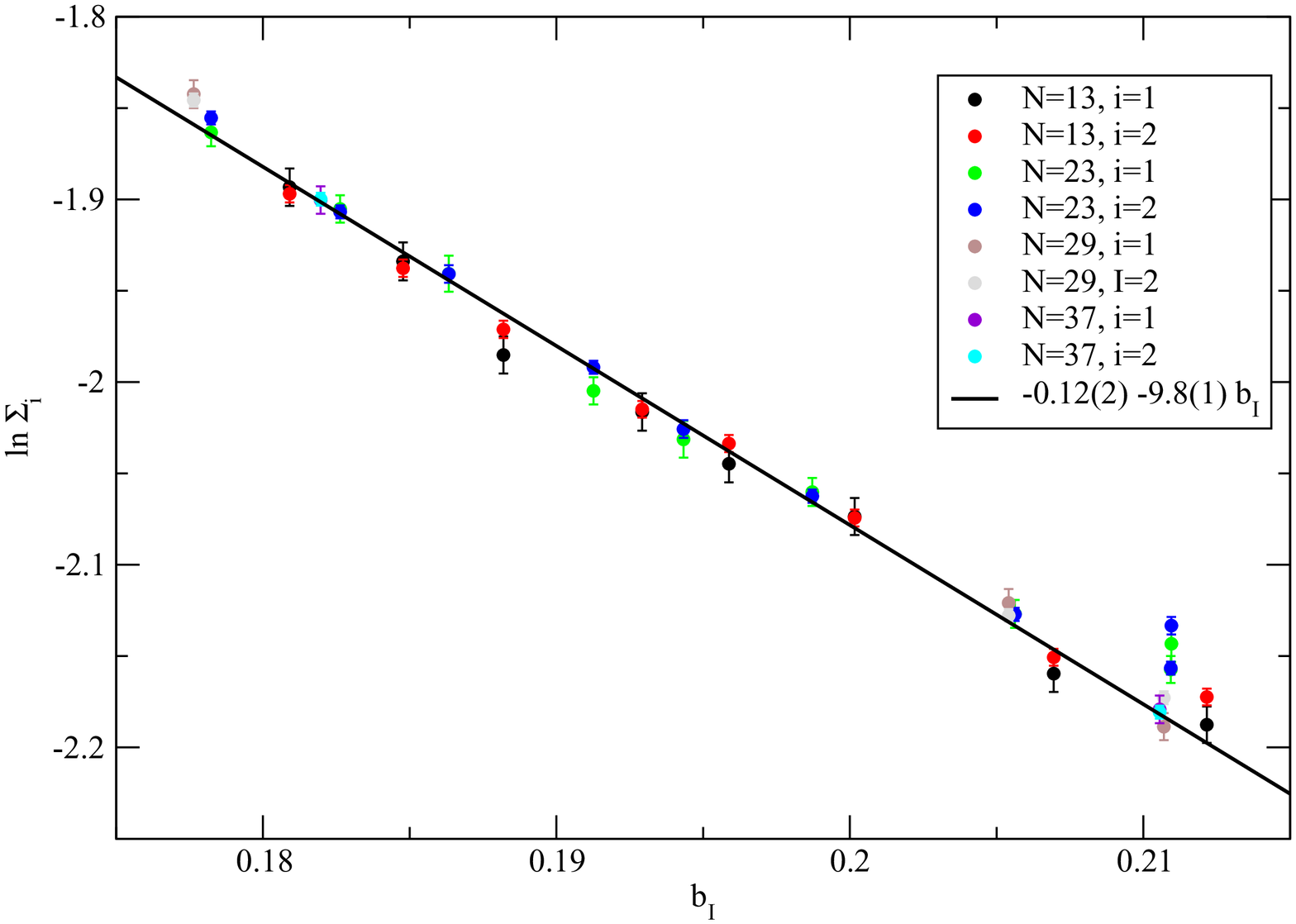}
\caption{ A plot of all the data for the condensate
listed in Table~\ref{tab3}. A linear regression fit is shown as a solid line.}
}

\subsection{The necessity of a nonabelian Thirring interaction in four dimensions}

We now proceed to investigate the ultraviolet problem in the four dimensional nonabelian case, in the limit of infinite $N$. The
fermions are restricted to the infinite $1-2$ plane in four dimensional
Euclidean space. The setup is a direct generalization of the 
three dimensional case. Working on an $L^4$ lattice 
we again find for all $b<b_c(L)$ that the lattice chiral symmetry is spontaneously
broken and chiral random matrix theory agrees with the numerical data.
Consequently, we can extract the condensate as a function of $b$.

Our results for the chiral condensate are summarized in Table~\ref{tab3}.

We have quoted the logarithm of the condensate since we expect it to
scale linearly with the coupling. Indeed a plot of the logarithm 
of the condensate as a function of the
tadpole improved coupling, $b_I=e b$, in Fig.~\ref{fig4} does show
a linear behavior. The one loop beta function for the four dimensional
theory would suggest a slope of $-\frac{24\pi^2}{11}$ and we obtain
a value that is roughly half of this number. This is in sharp distinction
with the situation in three dimensions. The exponential behavior might be
explained by an effective Nambu Jona-Lasinio model. 
In four dimensions the condensate
does not scale naively relative to bulk masses. 
The term $\LL_1 = J^j_\alpha J^j_\alpha$ from 
equation~(\ref{fourfermi}) indeed is needed, as perturbative analysis indicates.
With its introduction, 
a separate, adjustable, 
mechanism of dynamical scale generation will become operative, 
restricted to the plane, and the scaling of the condensate will be affected by it. 

The numerical implementation of the $\LL_1$ term with the required 
sign is a technical challenge.
The renormalization of the 2D/4D mixed system will require substantially more work.

We ignored taking the continuum limit for the time being and checked whether 
square cylinders in this case exhibited a large $N$ phase transition of
the same type as we have seen in three dimensions. Preliminary simulations
indicate that this is the case. Only after gaining control over the
continuum limit will we be able to address the question 
whether the structure of the large
$N$ phase transition is truly a feature of the continuum theory, as it is in 
three dimensions.

\section{Summary}

We have shown that Dirac fermions on a two dimensional surface embedded
in a three dimensional $SU(N)$ Yang Mills theory exhibit a large $N$ phase
transition separating short and large distance physics. The transition
occurs on a cylinder whose base has a scale $s$ when $s$ is equal to a
fixed number in units of the bulk correlation length. The exact 
value of the number will depend on the precise geometry of the cylinder
base. 

We have also seen that the $N$-dependence of the 
level spacing closest to zero goes from $\sim\frac{1}{N}$ in
the broken chiral symmetry phase to $\sim\frac{1}{N^{2/3}}$ in the symmetric
(gapped) phase. At the critical point, it goes as $\sim\frac{1}{N^{3/4}}$.
In this sense the situation is similar to that found for Wilson loops.
Also, the condensate vanishes at the critical size $l_c$ as $~\sim\sqrt{l-l_c}$.
Again, one can find an analogue of this in the Wilson loops case.

We suspect similar facts are 
true in four dimensions, except that there one extra free parameter will
have to be introduced in order to properly define the associated observable.
Thus, the critical size $l_c$ of the
base of a cylinder would depend on this extra parameter. 
We venture the guess that the exponents associated 
with $l-l_c$ and $N$ will not depend on this
extra parameter and will be the same as in 3D. 
Further work on the four dimensional case is needed.

\subsection*{Acknowledgments.}

R.N. acknowledges partial support by the NSF under grant number PHY-0854744.
HN acknowledges partial support by the DOE under grant
number DE-FG02-01ER41165.
HN notes with regret that his research has for a long time been 
deliberately obstructed by his high energy colleagues at Rutgers.   
HN wishes to thank Adam Schwimmer for many conversations. We thank Gernot
Ackemann, Poul Damgaard for Maciej Nowak for useful comments.

\end{document}